\documentclass[conference]{IEEEtran}
\IEEEoverridecommandlockouts
\usepackage{cite}
\usepackage{amsmath,amssymb,amsfonts}
\usepackage{algorithmic}
\usepackage{graphicx}
\usepackage{textcomp}
\usepackage{xcolor}
\def\BibTeX{{\rm B\kern-.05em{\sc i\kern-.025em b}\kern-.08em
    T\kern-.1667em\lower.7ex\hbox{E}\kern-.125emX}}

\graphicspath{{figures/}}
\usepackage{hyperref}
\usepackage[]{units}
\usepackage{amsthm}
\theoremstyle{definition}
\newtheorem{definition}{Definition}
\usepackage{booktabs}
\usepackage{color, soul}
\usepackage{url}
\usepackage{subcaption}
\usepackage{graphicx}
\usepackage{capt-of}
\usepackage[flushleft]{threeparttable}
\usepackage[T1]{fontenc}
\usepackage[utf8]{inputenc}
\usepackage[left]{lineno}
\usepackage[export]{adjustbox}
\usepackage{lipsum,multicol}
\usepackage{array}
\usepackage[super]{nth}
\newcolumntype{L}{>{\centering\arraybackslash}m{3cm}}
\newcolumntype{S}{>{\centering\arraybackslash}m{2.2cm}}
\newcolumntype{V}{>{\centering\arraybackslash}m{1.7cm}}

\begin{document}

\urldef\urlchem\url{https://www.statista.com/outlook/374/117/online-food-delivery/china#market-arpu}

\title{Exploring the Usage of Online Food Delivery Data for Intra-Urban Job and Housing Mobility Detection and Characterization}

\makeatletter
\newcommand{\linebreakand}{%
\end{@IEEEauthorhalign}
\hfill\mbox{}\par
\mbox{}\hfill\begin{@IEEEauthorhalign}
}
\makeatother


\author{\IEEEauthorblockN{Yawen Zhang\IEEEauthorrefmark{1},
		Seth Spielman\IEEEauthorrefmark{1}, Qi Liu\IEEEauthorrefmark{2}, Si Shen\IEEEauthorrefmark{1}, Jason Shuo Zhang\IEEEauthorrefmark{1}, Qin Lv\IEEEauthorrefmark{1}}
	\IEEEauthorblockA{\IEEEauthorrefmark{1} University of Colorado Boulder, Boulder, CO, USA\\ \IEEEauthorrefmark{2} Unsupervised Inc., Boulder, CO, USA\\
		Email: \IEEEauthorrefmark{1}\{yawen.zhang, seth.spielman, si.shen, jasonzhang, qin.lv\}@colorado.edu,
		\IEEEauthorrefmark{2}qliu.hit@gmail.com}}

\maketitle

\fussy
\begin{abstract}

Human mobility plays a critical role in urban planning and policy-making. However, at certain spatial and temporal resolutions, it is very challenging to track, for example, job and housing mobility. In this study, we explore the usage of a new modality of dataset, online food delivery data, to detect job and housing mobility. By leveraging millions of meal orders from a popular online food ordering and delivery service in Beijing, China, we are able to detect job and housing moves at much higher spatial and temporal resolutions than using traditional data sources. Popular moving seasons and origins/destinations can be well identified. More importantly, we match the detected moves to both macro- and micro-level factors so as to characterize job and housing dynamics. Our findings suggest that commuting distance is a major factor for job and housing mobility. We also observe that: (1) For home movers, there is a trade-off between lower housing cost and shorter commuting distance given the urban spatial structure; (2) For job hoppers, those who frequently work overtime are more likely to reduce their working hours by switching jobs. While this new modality of dataset has its limitations, we believe that ensemble approaches would be promising, where a mash-up of multiple datasets with different characteristic limitations can provide a more comprehensive picture of job and housing dynamics. Our work demonstrates the effectiveness of utilizing food delivery data to detect and analyze job and housing mobility, and contributes to realizing the full potential of ensemble-based approaches. 

\end{abstract} 

\begin{IEEEkeywords}
online food delivery; location profiling; job and housing mobility
\end{IEEEkeywords}

\section{Introduction}

\begin{figure}[t]
	\includegraphics[width=8.6cm]{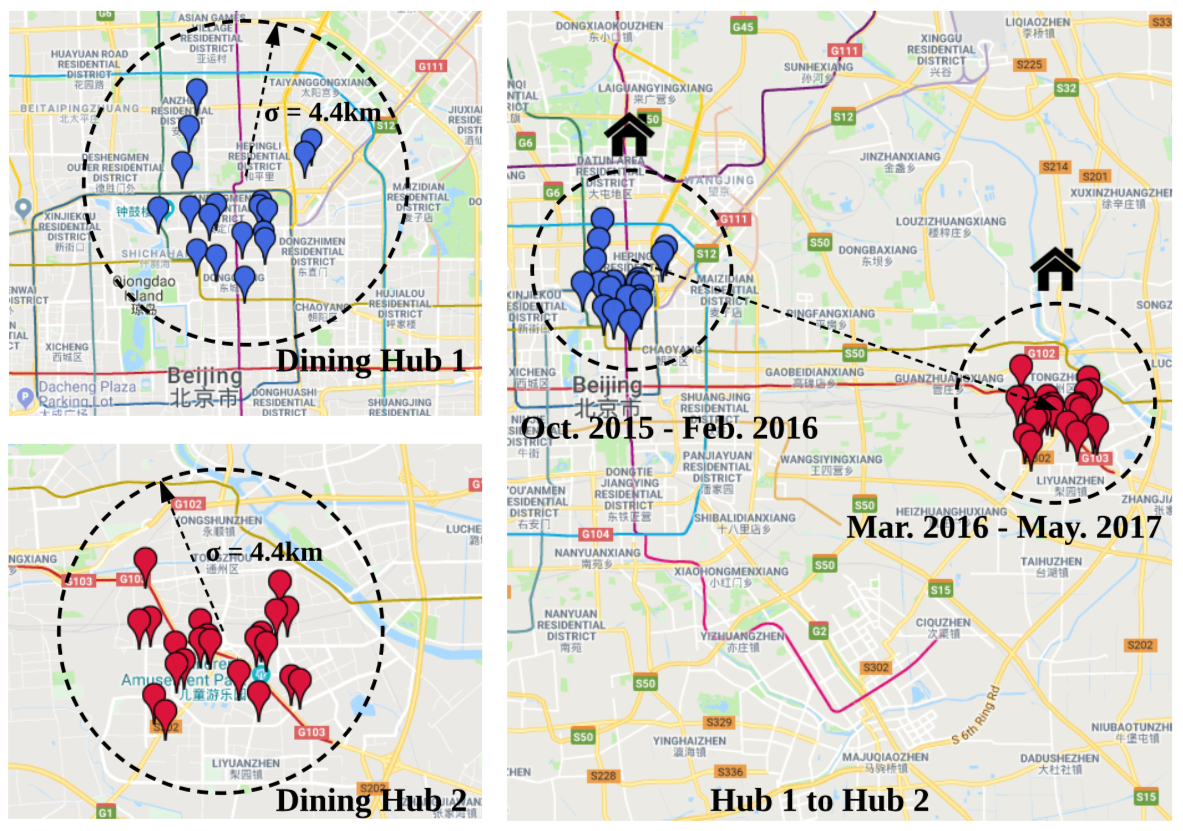}
	\caption{Examples of housing moves detected from food delivery service data, in which a user moved from dining hub 1 to 2 ($\sigma$: radius of a hub). Both hubs are the user's home hubs, which have no temporal overlapping. The colored dots represent the restaurants where the user ordered food. }
	\label{fig:transition}
\end{figure}

Human mobility, the movement of people in a city, is a central component for urban studies. It is multi-dimensional and could be understood at diverse temporal and spatial scales. At one extreme case, mobility can be observed at the time scale of seconds, such as movements of vehicles or pedestrians passing through road intersections. At another extreme, one might study life course events such as migrations using large-scale studies like the Panel Study of Income Dynamics (PSID). Human mobility is shaped by the surrounding environment (e.g., urban infrastructure), social factors (e.g., segregation and individual preferences), and economic constraints (e.g., cost of travel and housing). Mobility's inter-relationship with these forces makes it a strong indicator for understanding populations and managing urban space. However, at certain time scales, movements are challenging to detect, such as meso-scale job and housing mobility. 

The relationship between where one lives and works and a myriad of socioeconomic forces have led to an impressive body of literature on job and housing mobility~\cite{boyle2014exploring,coulter2016re}. Traditionally, to study this problem, researchers have heavily relied on survey and census data~\cite{cui2015residential,huang2017rural}. However, collecting these data is expensive. The activity of switching job or moving home is usually rare, thus difficult to capture in surveys based on random samples. In addition, due to infrequent updates in national census, mobility patterns extracted from such data are coarse-grained in a spatio-temporal sense, for example, highlighting only county-to-county moves on an annual basis. Furthermore, questions like ``Did you live in the house or apartment one year ago?'', which is asked on the largest US national survey, by design cannot capture frequent moves by an individual or household. Administrative records, like tax documents in the US, are reportedly limited in their ability to capture multiple moves within a single year. 

To observe finer-grained job and housing mobility, we look into new forms of data which enable the study of job and housing mobility. At aggregation level, internet search data have been explored to examine domestic migration~\cite{counts2019forecasting}. At individual level, Huang \textit{et al.} use smartcard data to track commuters' job and housing dynamics in Beijing~\cite{Huang12710}. As pointed out in~\cite{hughes2016inferring}, there is no ideal dataset and slightly different estimations may be derived from different data sources. However, different data sources could add complementary dimensions to the study of migrations. In this study, we explore the usage of online food delivery data to analyze job and housing dynamics. 


In recent years, food delivery service has gained significant popularity globally and has the potential for robust growth. It is significantly popular in China, about 50\% of Internet users (406 million) have used this service in 2018. Although used by a large population, it cannot be generalized to the whole population given the fact that most food delivery users are between 25 to 34 years old. However, it is important to point out that with food delivery data, we are looking at the most mobile group (people who are most likely to move) of ages between 25 and 35~\cite{doi:10.1080/02673030600807506}.

A key characteristic of food delivery service data is that, if a user regularly orders food from certain location, that place is very likely to be the user's home or workplace. Home and work locations are the two most frequent delivery locations for orders. The data that we have available do not expose a user's exact location, only the locations of restaurants. However, food delivery is constrained spatially, i.e., restaurants usually have a fixed delivery zone as shown in Figure~\ref{fig:transition}. We can then estimate a user's job and housing locations based on that user's restaurant orders because of these spatial constraints. We are not estimating users' exact addresses, but their approximate locations. The proposed estimation approach has some conceptual overlap with Backstrom {\em et al.}, which demonstrates that a user's physical location could be predicted by the known locations of his or her Facebook friends~\cite{backstrom2010find}. Because restaurants have delivery distance constraints and most of them only serve a limited geographic area, the inferences about users' home and work locations are well supported by this form of data.

In this work, we explore the usage of food delivery data to detect and characterize job and housing mobility, with case study in Beijing, China. Our main contributions are as follows: 
\begin{itemize}
    \item We propose a computational framework to infer users' home/work locations and detect job and housing moves using food delivery data. 
    \item Based on the job and housing moves detected, we analyze fine-grained job and housing spatio-temporal dynamics. 
    \item We further analyze various macro- and micro-level factors that are related to job and housing moves. 
\end{itemize}

\section{Related Work}

\subsection{Location Profiling}
With social media data, a user's home location can be inferred from his or her social network (friends) and user-centric data (tweets)~\cite{backstrom2010find,Twitter12}. Li {\em et al.} introduce a unified discriminative influence model that integrates locations observed from a Twitter user's friends, followers and tweets, which profiles the user's location effectively~\cite{Twitter12}. However, locations inferred from social media data are usually coarse-grained, e.g., city-level. Isaacman {\em et al.} infer home and work locations using cellular network data using a supervised method which achieves an accuracy of 88\% within 3 \unit{miles} of ground truth~\cite{pervasive11}. However, the performance of supervised method is strongly limited by the size of training data, in~\cite{pervasive11} only 18 volunteers's home and work locations are used. For large-scale online data, user locations are usually anonymized thus unsupervised methods are needed to infer locations. Previous studies~\cite{kung2014exploring, yang2018stay, Huang12710} dealing with large-scale data usually use rule-based identifications which define typical home and work patterns in advance. The approach we employ to identify home and work locations is similar to them except for using a clustering method to learn home and work characteristics from the data. 


\subsection{Human Mobility}
A variety of data sources have been used to study human mobility at different spatio-temporal scales. Previous studies have used social media data~\cite{ICWSM112831}, transportation data~\cite{yuan2013reconstructing} and mobile phone records~\cite{kung2014exploring} to capture short-term human mobility, which includes things that occur during the course of a few hours or a day, such as flows of people in the city. Compared with short-term mobility, job and housing mobility are ``slower'' and requires successive observations. There have been studies using Internet search data~\cite{counts2019forecasting} and transit smartcard data~\cite{Huang12710} to track aggregated and individual job and housing mobility. Our work is most related to them in exploring the usage of new forms of data to analyze job and housing dynamics. Without ground-truth data, it is important to use multiple perspectives to understand the same socioeconomic problem. Our study provides a complementary perspective into job and housing mobility. 

\subsection{Understanding Job and Housing Mobility.}
Previous studies on job and housing mobility mostly focus on understanding the decision making process, i.e., what are the major determinants of job and housing moves? Job and housing mobility are influenced by both macro- and micro-level factors. At macro-level, they are shaped by labor market~\cite{farber1999mobility, kim2014residential}, housing market~\cite{chan2001spatial} and urban zones (e.g., city core vs. suburbs)~\cite{liao2015changing}. At micro-level, an individual's job moving decision can be influenced by income, working conditions~\cite{tian2015investigating}, career opportunities~\cite{kronenberg2012move}, while housing moving decision could be influenced by residential satisfaction~\cite{speare1974residential}, commuting distance~\cite{clark2003does}, social relationship~\cite{musterd2016adaptive} and environmental factors (e.g., air quality)~\cite{chen2017effect}. However, there is a lack of studies on the large-scale spatiotemporal dynamics of job and housing mobility due to the limitations in traditional data sources. In this study, besides looking into the major factors related to job and housing mobility, we also leverage this new form of data to analyze the spatiotemporal characteristics of job and housing mobility. 
\section{Preliminaries}
\label{sec:problem} 

In this section, we describe definitions used in Section~\ref{sec:pre} and multi-source data that we collected. 

\subsection{Definitions}

\begin{definition}{Food Delivery Order. }
	The dataset consists of a list 
	of food delivery orders: 
	\[ \mathcal{O} = \{(u, r, t_d, t_c)\}, \] 
	where each tuple $(u, r, t_d, t_c)$ represents an order by user $u$ from restaurant $r$, the order was delivered at time $t_d$, and the 
	delivery took $t_c$ minutes. 
\end{definition}

\begin{definition}{Dining Hub.}
	\label{def:hub}
	For user $u$, given all the orders $\mathcal{O}_u$ made by $u$, we want to detect the user's dining hubs:  
	\[ \mathcal{B}_u = \{B_{u, 1}, B_{u, 2}, \ldots, B_{u, k}\}, \]
	where each hub contains a list of restaurants that the user ordered food from 
	and the corresponding orders. Note that the number of dining hubs $k$ can vary among users, and for a user with more than one dining hub, his/her dining hubs would be non-overlapping in terms of restaurants and food orders: 
	\[ \mathcal{O}_u = \bigcup_{i=1}^k \mathcal{O}_{B_{u,i}} \ \ \ and \] 
	\[ \mathcal{O}_{B_{u,i}} \cap \mathcal{O}_{B_{u,j}} = \emptyset, \ \ for\ i,j \in [1, k],  i\neq j\] 
\end{definition}

\begin{definition}{Hub Transition. }
	\label{def:hub_transition}
	Let $B_{u,i}$ and $B_{u,j}$ be two dining hubs of $u$, if $\mathcal{O}_{B_{u,i}}$ and 
	$\mathcal{O}_{B_{u,j}}$ do not overlap in time (i.e., orders of one dining hub all occurred
	before another dining hub's orders), and the two hubs' centers are certain distance apart, we determine there is a hub transition between $B_{u,i}$ and $B_{u,j}$. 
\end{definition}

\subsection{Multi-Source Data}

The food delivery dataset is crawled from Baidu Waimai food ordering website in 2017 (now acquired by eleme)~\footnote{The dataset and code used in this study are available from the first author (Yawen Zhang) upon request. }. This analysis uses a complete snapshot of Baidu Waimai's rating data during the period from June 26, 2014 to June 1, 2017 in Beijing, China. Each rating includes information about user, restaurant, delivery, and order ratings. The restaurants distribute densely in the study region. 


\begin{table}[h]
    \centering
	\scriptsize
	\caption{Selected attributes in each food delivery order. }
	\label{tb:ratingattr}
	\begin{tabular}{c|c}
		\hline
		\textbf{Attribute}  & \textbf{Description}  						\\ \hline \hline 
		User\_ID            & unique identifier of a user, anonymized for privacy  \\ \hline
		Restaurant\_ID      & unique identifier of a restaurant providing food-delivery \\ \hline
		Restaurant\_Lat/Lon & geolocation of the restaurant                             \\ \hline
		Arrive\_Time        & date and time when an order is delivered                  \\ \hline
		Cost\_Time          & time costed by food delivery  (minutes)                 \\ \hline
	\end{tabular}
\end{table}

The original Baidu Waimai dataset includes about 2.93 million users. Based on statistical analysis of the dataset, we find that the number of posted orders by a user follows the \textit{Pareto Principle}, i.e., a small portion of users contribute to the majority of posted orders. In order to have sufficient number of data points to detect hubs and hub transitions, we exclude users with less than 10 orders posted online. After filtering out ad-hoc users, this dataset contains nearly 27 million food delivery orders generated by about 0.7 million users. 





To contextualize job and housing moves and show how they inform our understanding of broader economic trends, we collected supplemental data from the latest nationwide census as well as public websites. 

\begin{itemize} 
	\item  {\bf Employment statistics:} The employment statistics at subdistrict level are obtained 
	from the third National Economic Census (till 2014) in China. 
	\item {\bf Population statistics:} The residential population statistics at subdistrict level are obtained 
	from the sixth National Population Census (till 2010). 
	\item {\bf Housing price:} Monthly house transaction data (with price and location) from June 2014 to June 2017 in Beijing are collected from NetEase. 
\end{itemize} 




\section{Computational Framework}
\label{sec:pre} 

We propose a two-step approach to detect job and housing moves from food delivery data, which involves (1) dining hub detection and (2) job and housing moves detection. 

\subsection{Dining Hub Detection}

Since users' locations are unknown, we propose a weighted kernelized MeanShift (WKMS) clustering method to detect each user's dining hubs as described in Definition~\ref{def:hub}.

\paragraph{WKMS Clustering Approach}
The task is to detect the dining hubs $\mathcal{B}_u$ of user $u$, given $u$'s historical food delivery orders $\mathcal{O}_u$. Since users typically order food from a few fixed locations (e.g., home or workplace), and restaurants usually deliver within certain distances, we can identify $u$'s dining hubs based on the groups of adjacent restaurants the user ordered food from, and restaurants belonging to different dining hubs are usually farther away from each other compared with those in the same hub. Hence, the dining hub detection problem can be formulated as a clustering problem. 

The intuition behind our design is two-fold.  
(i) Although the number of dining hubs belonging to a user is unknown, the user's hub centers are stationary points that can be estimated by the underlying probability density function regarding the user's dining preferences. Thus, a non-parametric density function estimation approach with minimal initializing parameters is preferred. 
(ii) Given the average food delivery time from a restaurant $r$ to a user $u$, a shorter delivery time indicates 
that this restaurant is closer to the user's hub center. Thus, the average food delivery time is a proper variable to estimate the distance from a restaurant to the hub center or the user's actual stay point. 

Specifically, WKMS clustering method adopts a weighted Gaussian kernel as defined in Equation~\ref{eqn:wkms_mean} and \ref{eqn:wkms_kernel},
\begin{equation}
\label{eqn:wkms_mean}
m(r^{i}) = \frac{\sum_{r^{j}\in N(r^{i})}K(r^{j}-r^{i})r^{j}}{\sum_{r^{j}\in N(r^{i})}K(r^{j}-r^{i})}, 
\end{equation}
\begin{equation}
\label{eqn:wkms_kernel}
K(r^{j}-r^{i}) = w_{r^{j}}e^{\frac{-||r^{j}-r^{i}||^2}{2\sigma^2}}, 
\end{equation}
where $R_u =\{r^1, r^2, ... r^{n}\}$ is the set of restaurants that user $u$ has ordered food from, $N(r^i)$ are the neighbors of restaurant $r^i$, $||r^{j}-r^{i}||$ is the distance between two restaurants, and $m(r^{i})$ is the mean shift vector for restaurant $r^{i}$.  
The weight $w_{r^{j}}$ is a function of the food delivery time: 
\begin{equation}
\label{eqn:wkms_weight}
w_{r^{j}} = \frac{1}{avg({r^j_{u, t_c}})},  
\end{equation}
where $avg({r^j_{u, t_c}})$ is the average delivery time $t_c$ among all orders made by user $u$ from restaurant $r^j$. 

WKMS is derived from MeanShift clustering~\cite{cheng1995mean} which is based on kernel density estimation (KDE). This clustering process has great similarity with food delivery process, in which a user's stay point represents the localized hotspot (with high density) for orders.  

\paragraph{Bandwidth Selection}

The only parameter in the WKMS method is the bandwidth parameter $\sigma$ (i.e., radius) in Equation~\ref{eqn:wkms_kernel}. A spatial constraint is given by $\sigma$ as: 
\begin{equation}
	d(r_u^j, B_{u, i}) \leq \sigma, 
\end{equation}
where $C(r_u^j) = B_{u, i}$, $C(r_u^j)$ is the cluster that $r_u^j$ belongs to. $d(r_u^j, B_{u, i})$ represents the distance between restaurant $r_u^j$ and estimated center of $B_{u, i}$. In food delivery scenario, $\sigma$ represents the average delivery distance of restaurants as shown in Figure \ref{fig:transition}.

We estimate $\sigma$ using Baidu Waimai queries. We conducted queries corresponding to 540 different locations evenly distributed in the study area in Beijing. For each query, Baidu Waimai would return all the available restaurants for delivery, as well as their delivery methods and distances. We thus gathered data from 20,294 restaurants to better understand the delivery distance constraints, which are used to estimate $\sigma$. 

According to the queries, food deliveries are carried out by four different methods, including baidulogistics, baiduzhongbao, cityexpress and self (delivery by restaurant itself). As shown in Table \ref{tb:log_type}, self delivery has longer delivery distance than the other methods but it is only offered by around 12\% of the restaurants. 

\begin{table}[h]
	\small
	\centering
	\caption{The percentages and delivery distances of different food delivery methods. }
	\label{tb:log_type}
	\begin{tabular}{c||c|S}
		\hline 
		\bf Delivery Method  & \bf Percentage (\%) & \bf \nth{95} Percentile Distance (km) \\
		\hline \hline 
		baidulogistics & 60.11 & 3.35    \\
		\hline
		baiduzhongbao  & 4.73 & 3.70    \\
		\hline
		cityexpress    & 23.57 & 4.18     \\
		\hline
		self           & 11.58 & 14.87   \\
		\hline
	\end{tabular}
\end{table}

The \nth{95} percentile delivery distance of all methods combined is around 4.4km, and the \nth{99} percentile delivery distance is around 13.1km, mainly contributed by the self delivery cases. It should be noted that longer delivery distance orders require an extra delivery fee, which does not apply for most delivery orders. Hence, for the general clustering purpose, we set the bandwidth $\sigma$ to 4.4km by considering the majority of food ordering cases. 

\subsection{Job and Housing Moves Detection} \label{rule}

\paragraph{Home and Work Hubs Identification} Though $H$ (Home) and $W$ (Work) hubs are the two most frequent places for food ordering, there could also be some temporary hubs, e.g., visiting a friend's home. It is important to filter out temporary hubs before identifying home and workplace. Specifically, we use hub frequency and duration as constraints: (i) hub frequency: the number of orders at a hub should be no less than 10\% of total orders of a user, (ii) hub duration: a user should continuously order food from a hub for no less than 30 days. These thresholds are set by experiments. 

We employ a similar approach used in~\cite{kung2014exploring, yang2018stay, Huang12710} to identify $H$ and $W$ hubs. In detail, for each hub, we compute its ordering frequencies during multiple time slots, including morning (6am--11am), noon (11am--3pm), afternoon (3pm--7pm), evening (7pm--10pm) and night (10pm--6am) of weekday, weekends and holiday. K-means clustering is used to form clusters based on these temporal features. By experimenting with different number of clusters to maximize the Silhouette score, we set n\_clusters to 4. We can identify clear home and work clusters, in which $W$ hubs mostly order food during weekday noon and rarely on weekends or holidays, while $H$ hubs mostly order food during weekends or holidays and rarely on weekdays. By examining their spatial locations, we observe that most $W$ hubs are located in business or industrial areas while $H$ hubs are in residential areas. 


\paragraph{Job and Housing Moves Detection}

With Definition~\ref{def:hub_transition}, we detect $H \rightarrow H$ as housing move and $W \rightarrow W$ as job move. Similar to the group definitions in Huang {\em et al.}~\cite{Huang12710}, we identify three user groups: (1) Stayers, users without change in $H$ or $W$ hub; (2) Job hoppers, users who change $W$ hub; and (3) Home movers, users who change $H$ hub. 

\section{Findings and Insights}

\subsection{Statistics on Job and Housing Locations \& Moves}

In total, 852,572 hubs are detected from 584,563 users with the WKMS approach. About 21\% users are removed because their dining orders cannot form clusters. We identify 169,612 $H$ and 231,419 $W$ hubs from 293,620 users, and these users are used for job and housing mobility analysis.  





The identification of $H$ and $W$ hubs forms the basis for move detection. We conduct two validations on the detected $H$ and $W$ hubs: (1) Hotspots validation, as shown in Figure~\ref{fig:hub_distribution}, the identified hotspots for $H$ and $W$ hubs coincide well with the local household surveys conducted in Beijing. $H$ and $W$ hubs do not share common hotspots, and home hubs have more hotspots in the suburbs. It confirms the ``Home-Work Separation'' situation in Beijing~\cite{liu2009home}, i.e., a lot of people select home location different from work location. (2) Ratio comparison, at subdistrict-level, we compute the total number of $H$ and $W$ hubs within each subdistrict and compute its work-home-ratio (142 subdistricts in total). We then compare the job-housing-ratio derived from census data and work-home-ratio derived from food delivery data. The Pearson correlation coefficient between them is 0.74, which indicates that the detected $H$ and $W$ hubs have good consistency with census data. 


\begin{figure}[t]
	\begin{subfigure}[h]{0.23\textwidth}
		\centering
		\includegraphics[width=\textwidth]{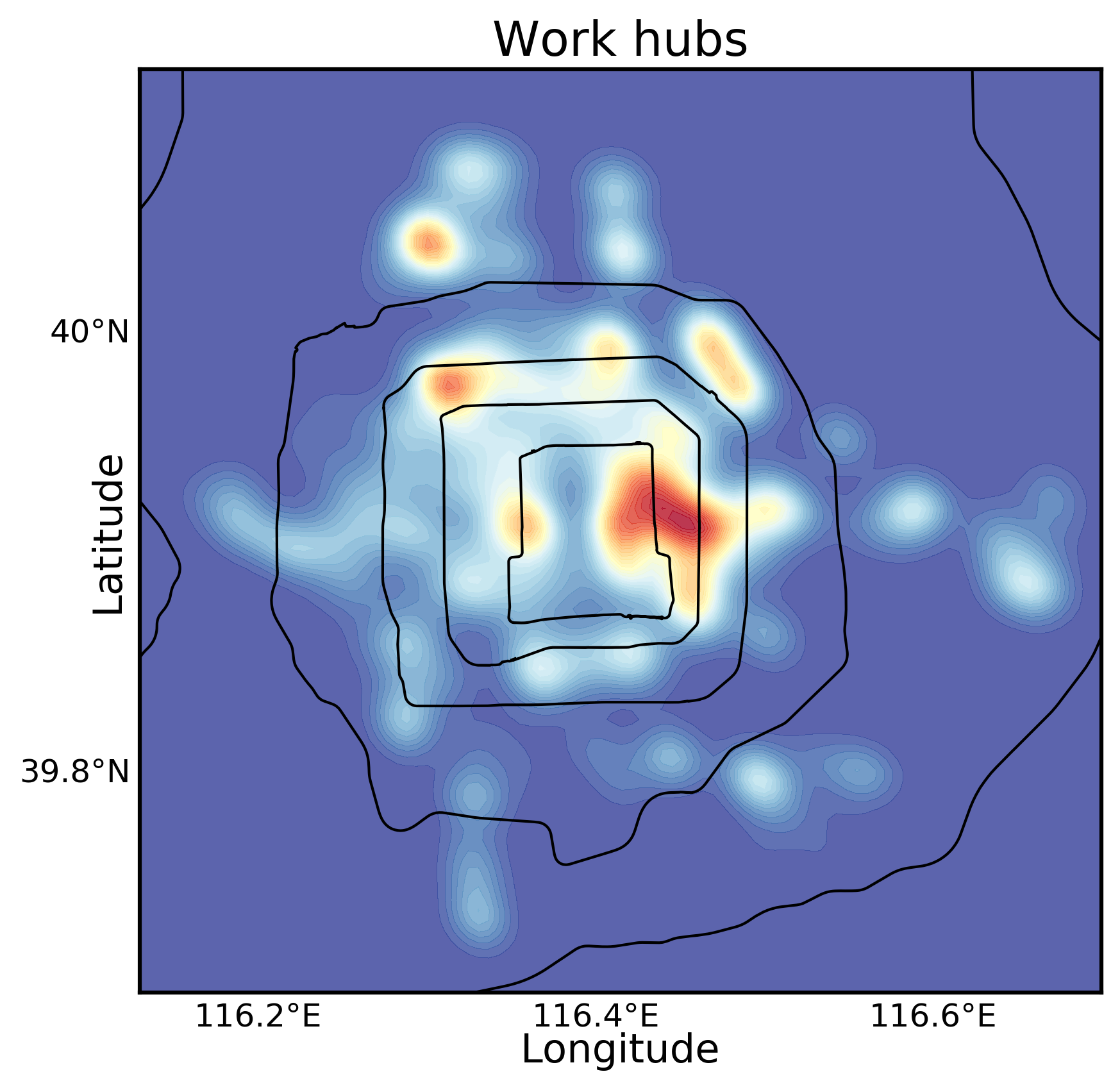}
		\caption{Work hubs}
		\label{fig:W_hubs}
	\end{subfigure}
	\begin{subfigure}[h]{0.225\textwidth}
		\includegraphics[width=\textwidth]{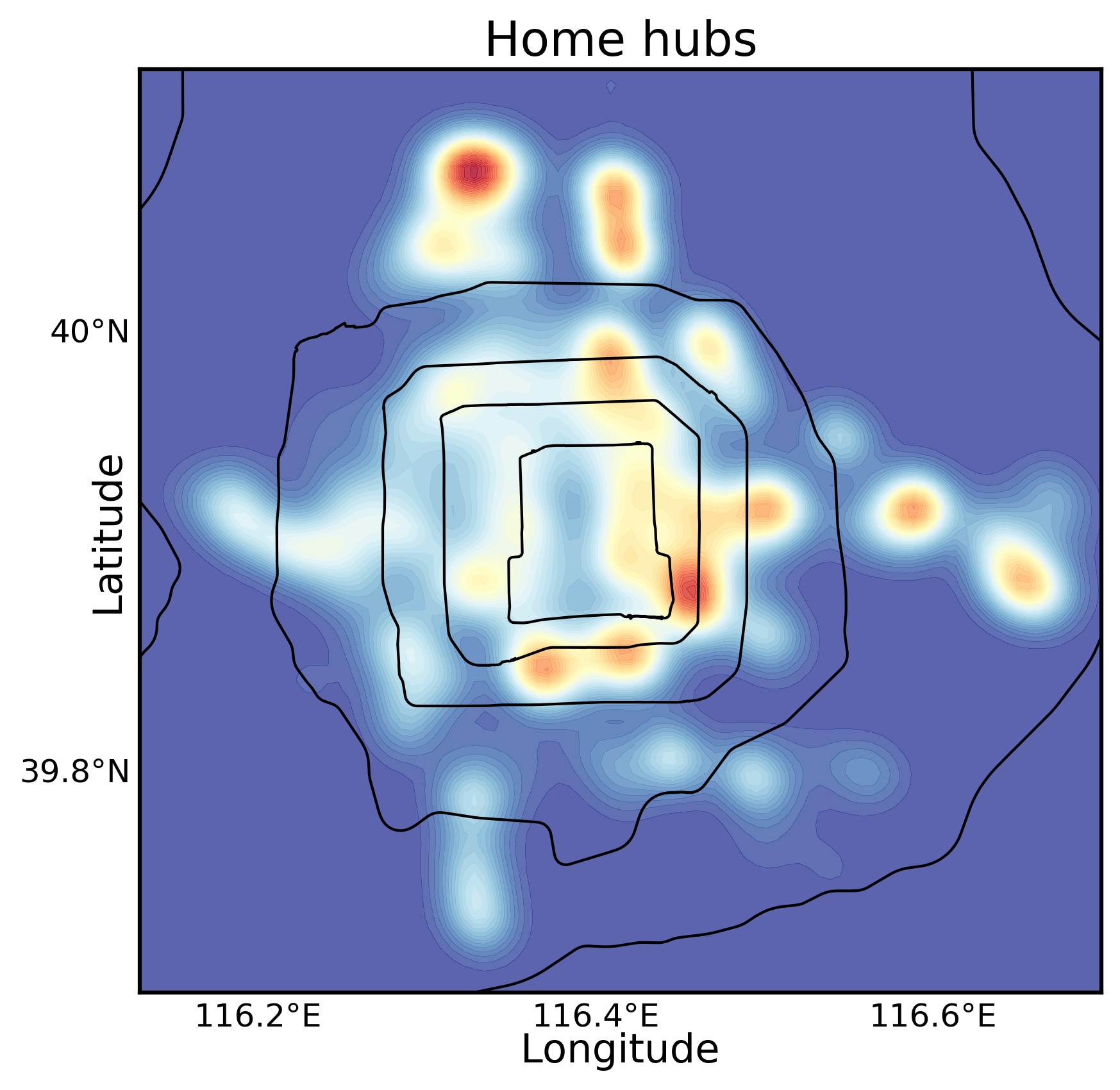}
		\caption{Home hubs}
		\label{fig:H_hubs}
	\end{subfigure}
	\caption{Spatial distributions of home and work hubs using 2D Kernel Density Estimation.} 
	\label{fig:hub_distribution}
\end{figure}


Based on the group definitions, we identify 17,249 stayers, 4,138 job hoppers, and 2,368 home movers. It should be noted that only a small percentage of users are detected as job hopper or home mover. Similarly, in~\cite{Huang12710}, among 5 million smartcard users only 4,248 regular commuters are selected. The study of job and housing mobility requires successive observations on workplace and residence over years. 


\subsection{Collective Job and Housing Dynamics}

From the macro-scale perspective, temporal and spatial characteristics in job and housing mobility are not well understood in Beijing with existing data sources. By aggregating job and housing moves, we analyze the collective patterns observed from food delivery data and engage in a kind of ``soft'' validation of the detected moves by comparing them with the literature on job and housing dynamics. 

\vspace{0.15cm}
\textit{\textbf{Observation 1:} There is a synchronized seasonal pattern in job and housing moves and people are more likely to move for targeted reasons such as employment opportunity.} 
\vspace{0.15cm}

We compute the total number of job and housing moves in each month. As shown in Figure~\ref{fig:W_H_month}, job and housing moves demonstrate synchronized seasonal patterns. Their temporal variations can be explained as follows: (1) The peak time in March corresponds to ``Gold March Silver April'' period in China, during which time a lot of people switch their jobs after receiving bonus of previous year. Due to the synergy between job and housing movement, it also causes a lot of people relocating their housing during this period. (2) Another peak in July and August corresponds to the large-scale job enrollment during summer, which results in a large number of job and housing moves. The temporal patterns are consistent with the general characteristics in Beijing, which provides validity to the detected job and housing moves. Moreover, the relative differences and seasonal patterns can be further examined to understand the exact causes behind those movements. 

\begin{figure}[t]
	\centering
	\includegraphics[width=6.5cm]{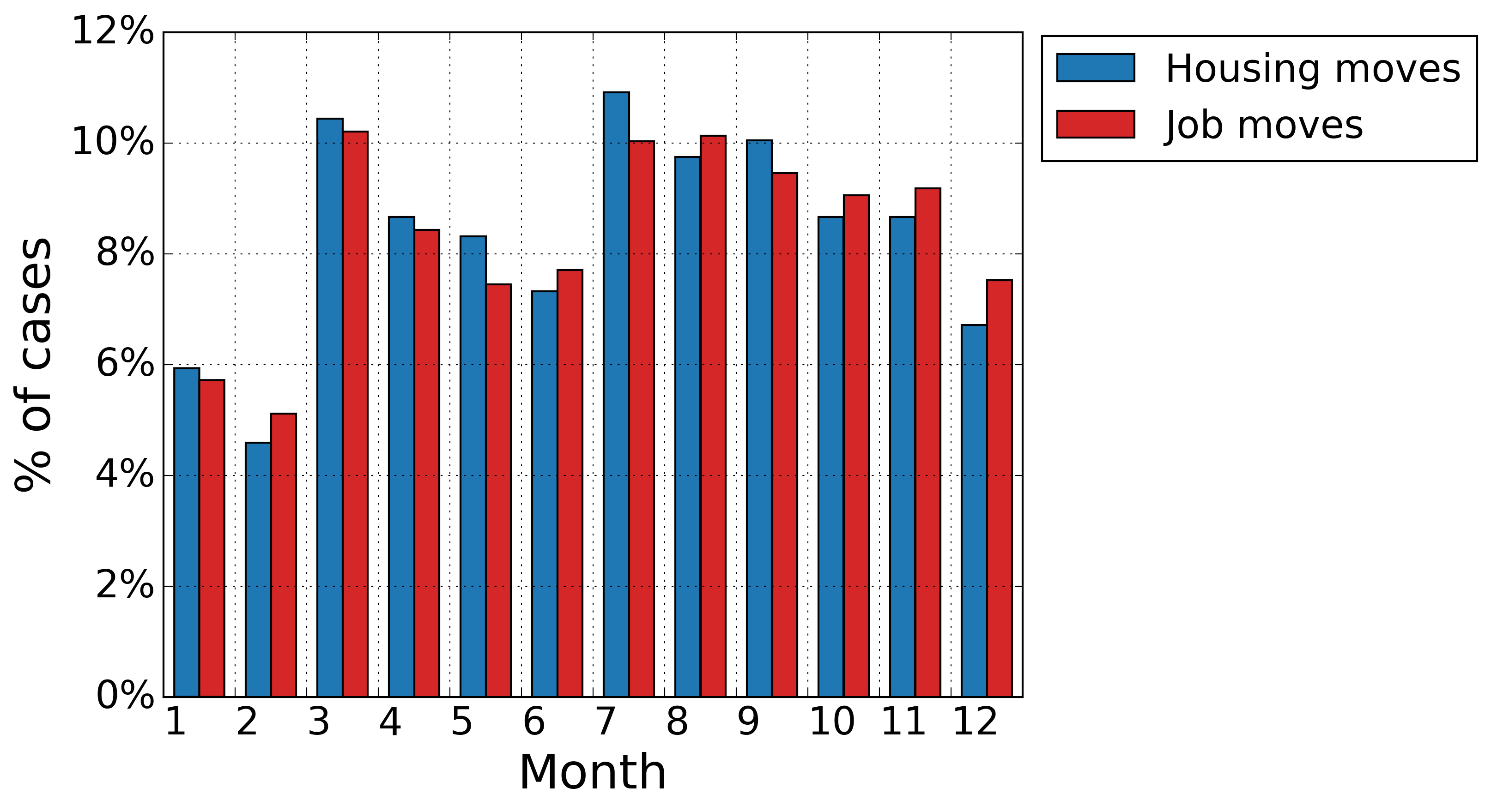}
	\caption{Temporal distribution of job and housing mobility.}
	\label{fig:W_H_month}
\end{figure}

\vspace{0.15cm}
\textit{\textbf{Observation 2:} In terms of spatial dynamics, job moves are more concentrated than housing moves and there are substantial housing moves into suburban of the city. } 
\vspace{0.15cm}

The study area has 142 subdistricts. We use them to summarize spatial trends in job and housing mobility. Subdistrict is the basic geographic unit in Chinese census that collects population and employment information. Specifically, we construct two directed graphs as shown in Figure \ref{fig:two_flow} based on the detected job and housing moves. 


\begin{figure}[t]
	\begin{subfigure}{0.23\textwidth}
		\includegraphics[width=\textwidth]{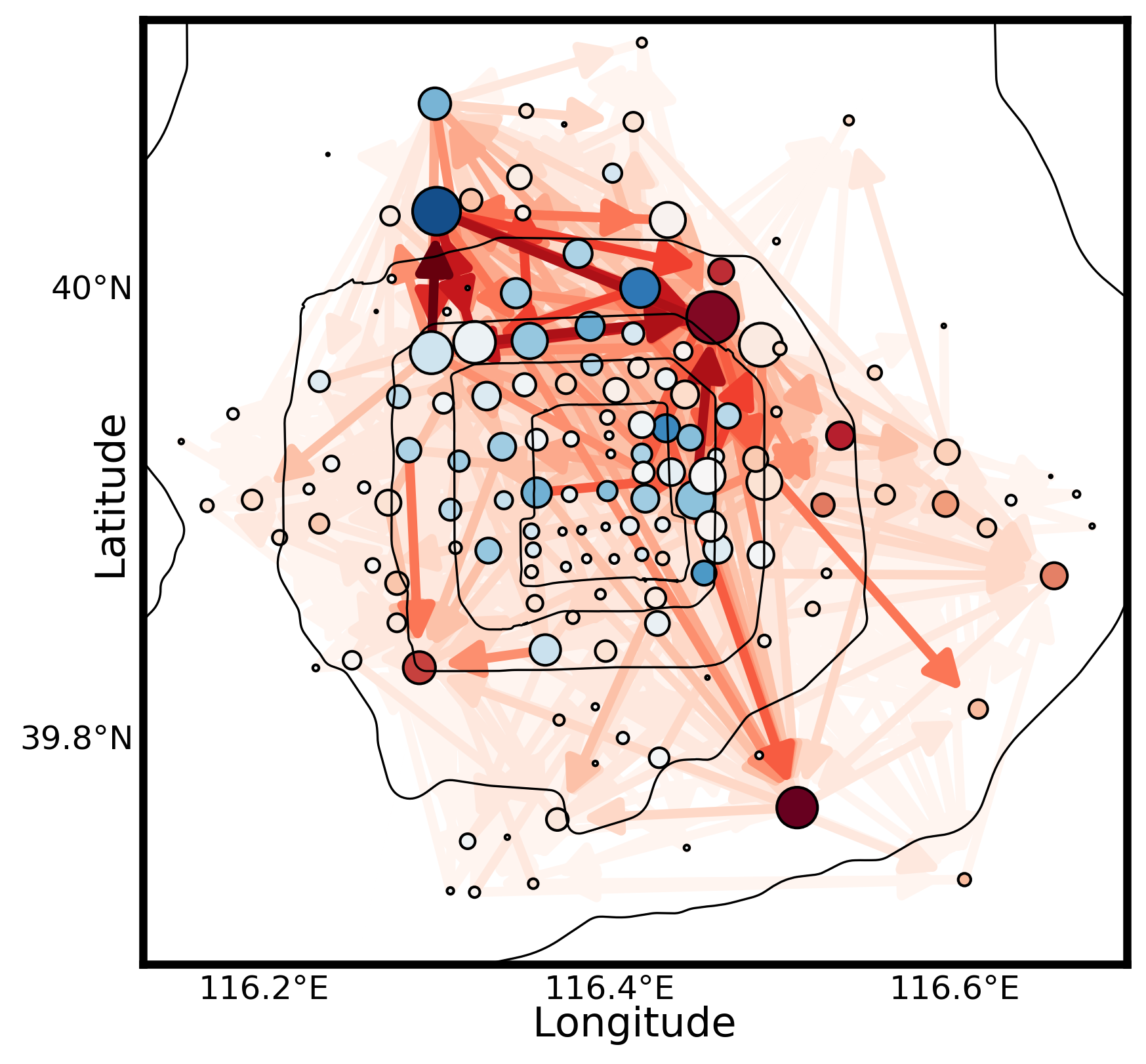}
		\caption{Job dynamics}
		\label{fig:job_flow}
	\end{subfigure}
	\begin{subfigure}{0.23\textwidth}
		\includegraphics[width=\textwidth]{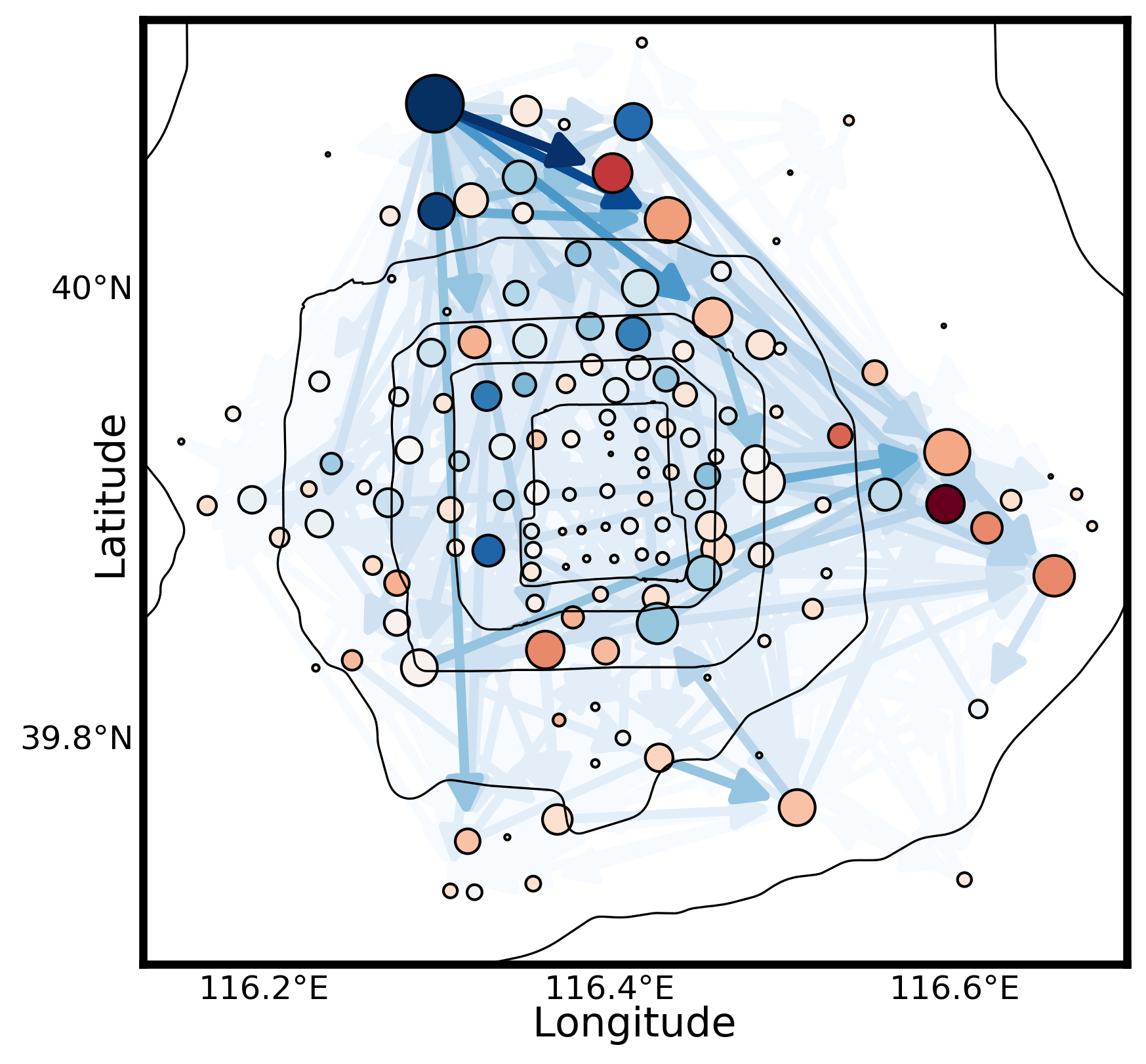}
		\caption{Housing dynamics}
		\label{fig:house_flow}
	\end{subfigure}
	\caption{Directed graphs of job and housing moves (Each node is a subdistrict. Its size represents the total number of moves in the subdistrict, the larger the more moves. Its color represents \textit{in} minus \textit{out} moves, red indicates more \textit{in} moves and blue indicates more \textit{out} moves. Each link is a moving direction, and its color represents its frequency, the darker the more frequent.)} 
	\label{fig:two_flow}
\end{figure}

In Figure~\ref{fig:two_flow}, job and housing moves demonstrate different spatial patterns. For job moves, most active regions are within the 4th Ring Road, i.e., city core, for example, subdistricts in \textit{Haidian} district with dominant function as Science/Education/Technology area~\cite{yuan2015discovering}. In Figure~\ref{fig:job_flow}, we can identify frequent moving trajectories, whose origins and destinations correspond to the popular industrial and business regions in Beijing. In comparison, housing moves are less concentrated in the central city. For housing moves, the majority of active subdistricts are outside central Beijing, which is consistent with the literature describing rapid suburban growth in Beijing~\cite{liu2018urban}. The most active subdistricts are located in North and East Beijing where there are more residential areas. Although the North and the East were both active, in terms of attractiveness, the East were attracting more people. The rapid growth of the East is due to Beijing Administrative Center's move to \textit{Tongzhou} district in the East by 2018~\cite{lin2017study}, which makes this area a popular residential destination. Compared with job moves, housing moves are diffused with less frequent moving trajectories, and a lot of people moved to suburban or peri-urban areas, such as \textit{Changping}, \textit{Fengtai}, \textit{Daxing} and \textit{Tongzhou} districts. 


\subsection{Individual Job and Housing Dynamics}

There are various factors influencing job and housing moving decisions. Although we have limited insights into individual moving decisions compared with longitudinal analysis, we can examine the emergent patterns in job and housing moves. In this section, we link individual job and housing moves with macro- and micro-level factors, including commuting distance, housing price, working behavior and urban spatial structure, in order to better understand the main motivations behind job and housing moves. 

\vspace{0.15cm}
\textit{\textbf{Observation 3:} A longer commuting distance encourages both job and housing mobility, and the commuting distance tolerance in Beijing is about \unit[11.75]{km}. } 
\vspace{0.15cm}

In mega-cities like Beijing, people usually have to commute a long distance from home to workplace~\cite{long2015combining}. With the dataset, if we detect both $H$ and $W$ hubs of a user, we can then compute the user's commuting distance. As shown in Table~\ref{tb:commute_distance}, for different user groups, we report their average commuting distances. The average commuting distance for stayers is \unit[11.75]{km} which is consistent with the reported average commuting distance for white-collar workers in Beijing. Job hoppers and home movers have longer commuting distances than stayers, which indicates that a longer commuting distance encourages both job and housing mobility. Also, the distributions of commuting distances shown in Figure~\ref{fig:commute_density} are consistent with the general home-work commuting patterns~\cite{kung2014exploring}. 


\begin{table}[h]
	\centering
	\caption{Comparison of different user groups on commuting distance (pre-move commuting distances are computed for job hoppers and home movers). }
	\small
	\begin{tabular}{V||S|S}
		\hline
		\textbf{User Group} & \textbf{Avg Commuting Distance (\unit{km})} & \textbf{t-statistic (Comparison vs. Stayer)}         \\ 
		\hline \hline 
		Stayers & 11.75 & /  \\ \hline	
		Job joppers  & 13.06 & 7.372***  \\ \hline
		Home movers  & 12.57 & 3.750***  \\ \hline
	\addlinespace[0.75ex]
    \multicolumn{2}{l}{\textsuperscript{***}$:p<.001$, 
    \textsuperscript{**}$:p<0.01$, 
    \textsuperscript{*}$:p<0.1$}
	\end{tabular}
	\label{tb:commute_distance}
\end{table}

For job hoppers and home movers, we also compare their pre-move and post-move commuting distances and find no significant difference, which indicates that the distribution of commuting distances tend to be stable. Some people decrease their commuting distances while others increase their commuting distances, i.e., a balance is struck. However, the story is different when we further examine users with different pre-move commuting distances. From Figure~\ref{fig:commute_difference}, we observe that people with longer pre-move commuting distances are more likely to reduce their commuting distance with job or housing moves. Home movers generally reduce more commuting distance than job hoppers. The commuting distance tolerance for both job and housing moves are approximate to the average commuting distance of stayers. 

\begin{figure}[t]
	\begin{subfigure}[h]{0.235\textwidth}
		\centering
		\includegraphics[width=\textwidth]{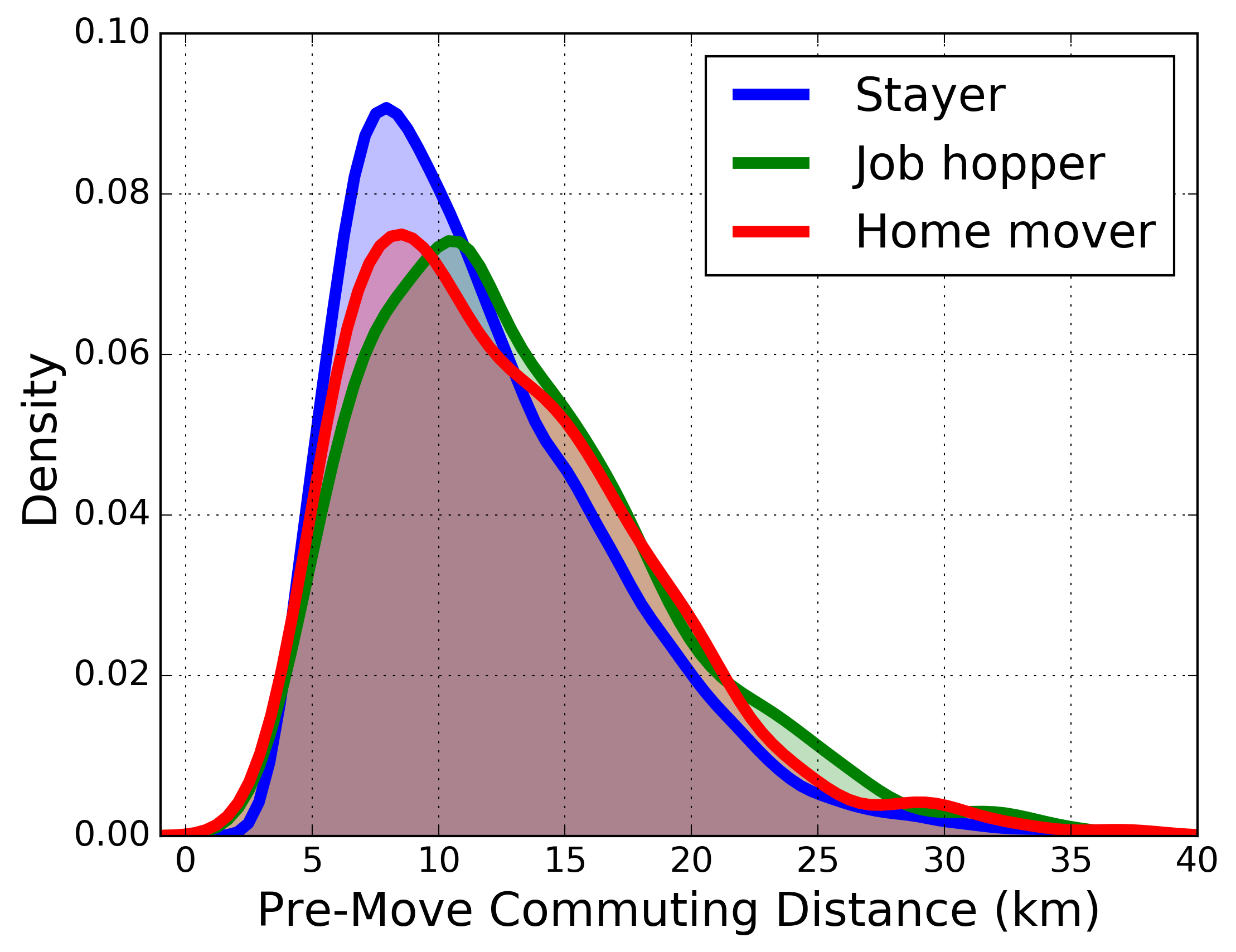}
		\caption{Density plot}
		\label{fig:commute_density}
	\end{subfigure}
	\begin{subfigure}[h]{0.23\textwidth}
		\includegraphics[width=\textwidth]{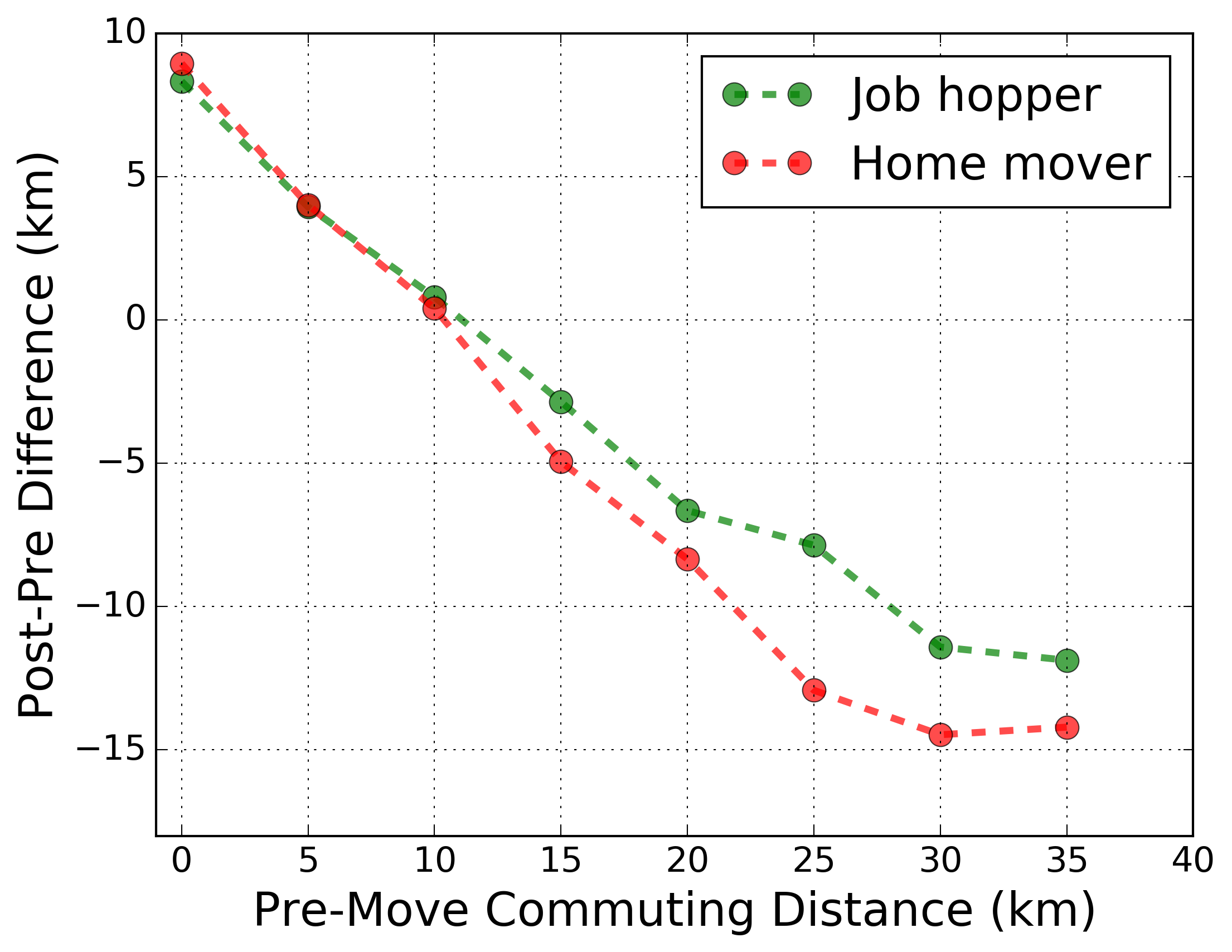}
		\caption{Post-Pre difference}
		\label{fig:commute_difference}
	\end{subfigure}
	\caption{a) Density plots of commuting distances for different user groups, b) Average post-pre commuting distance differences at different pre-move commuting distances (which are divided into \unit[5]{km} bins).}
	\label{fig:commnute}
\end{figure}

\vspace{0.15cm}
\textit{\textbf{Observation 4:} People who frequently work overtime are more likely to reduce their working time by job hopping. } 
\vspace{0.15cm}

With food delivery data, we are able to look into users' working behaviors, for example, whether a user frequently works overtime can be detected if the user usually orders food at a $W$ hub outside regular working hours, i.e., weekday nights or weekends. With this information, we can try to estimate the impacts of job hopping on users' working behaviors. Specifically, we compute the working overtime ratio as the relative frequency of food orders happening on weekday evening or night and anytime on weekends. 

We compare stayers and job hoppers on their average working overtime ratio. While job hoppers have slightly higher ratio, their difference is not significant as shown in Table~\ref{tb:work_time}. Job hoppers' post-move working overtime ratio is slightly lower than their pre-move ratio. The results indicate job hoppers are not necessarily different from stayers in terms of working overtime since there can be various factors influencing individual job hopping decisions. 

\begin{table}[h]
	\centering
	\caption{Comparison on working overtime ratio (Stayers: job hubs are used, Job hoppers: pre-move hubs are used). }
	\small
	\begin{tabular}{V||S|S}
		\hline
		\textbf{User Group} & \textbf{Avg Working Overtime Ratio}      & \textbf{t-statistic (Comparison vs. Stayer)}         \\ 
		\hline \hline 
		Stayers (job) & 0.207 & /  \\ \hline
		Job hoppers  & 0.209 & 0.262  \\ \hline		
	\addlinespace[0.75ex]
    \multicolumn{2}{l}{\textsuperscript{***}$:p<.001$, 
    \textsuperscript{**}$:p<0.01$, 
    \textsuperscript{*}$:p<0.1$}
	\end{tabular}
	\label{tb:work_time}
\end{table}

However, when comparing job hoppers with different pre-move working overtime ratios as shown in Figure~\ref{fig:work_difference}, we observe that with higher working overtime ratio, people are more likely to reduce their working overtime after job hopping. The dropping in ratio is more significant for users with extremely high working overtime ratio. 

\begin{figure}[t]
	\begin{subfigure}[h]{0.235\textwidth}
		\centering
		\includegraphics[width=\textwidth]{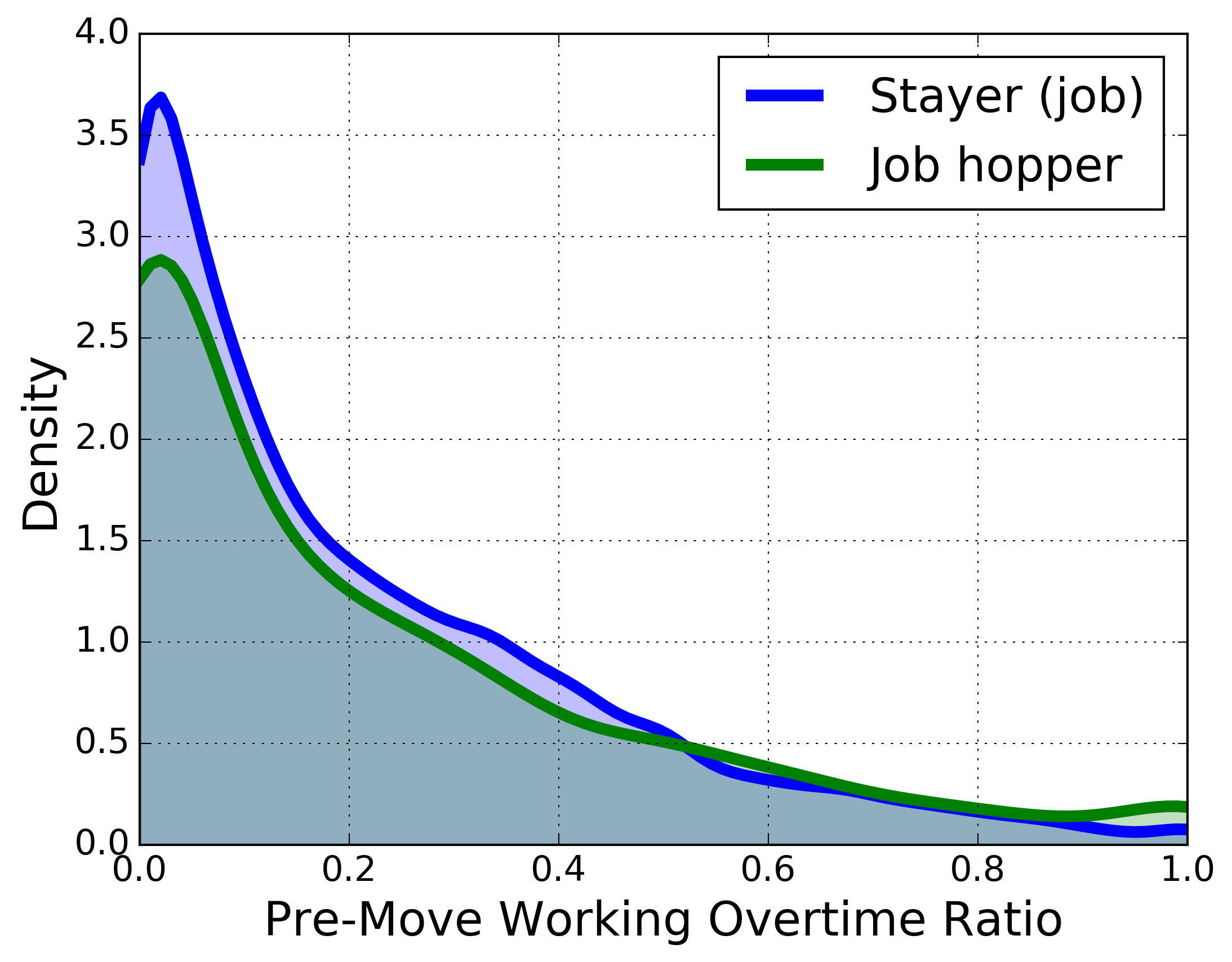}
		\caption{Density plot}
		\label{fig:work_density}
	\end{subfigure}
	\begin{subfigure}[h]{0.23\textwidth}
		\includegraphics[width=\textwidth]{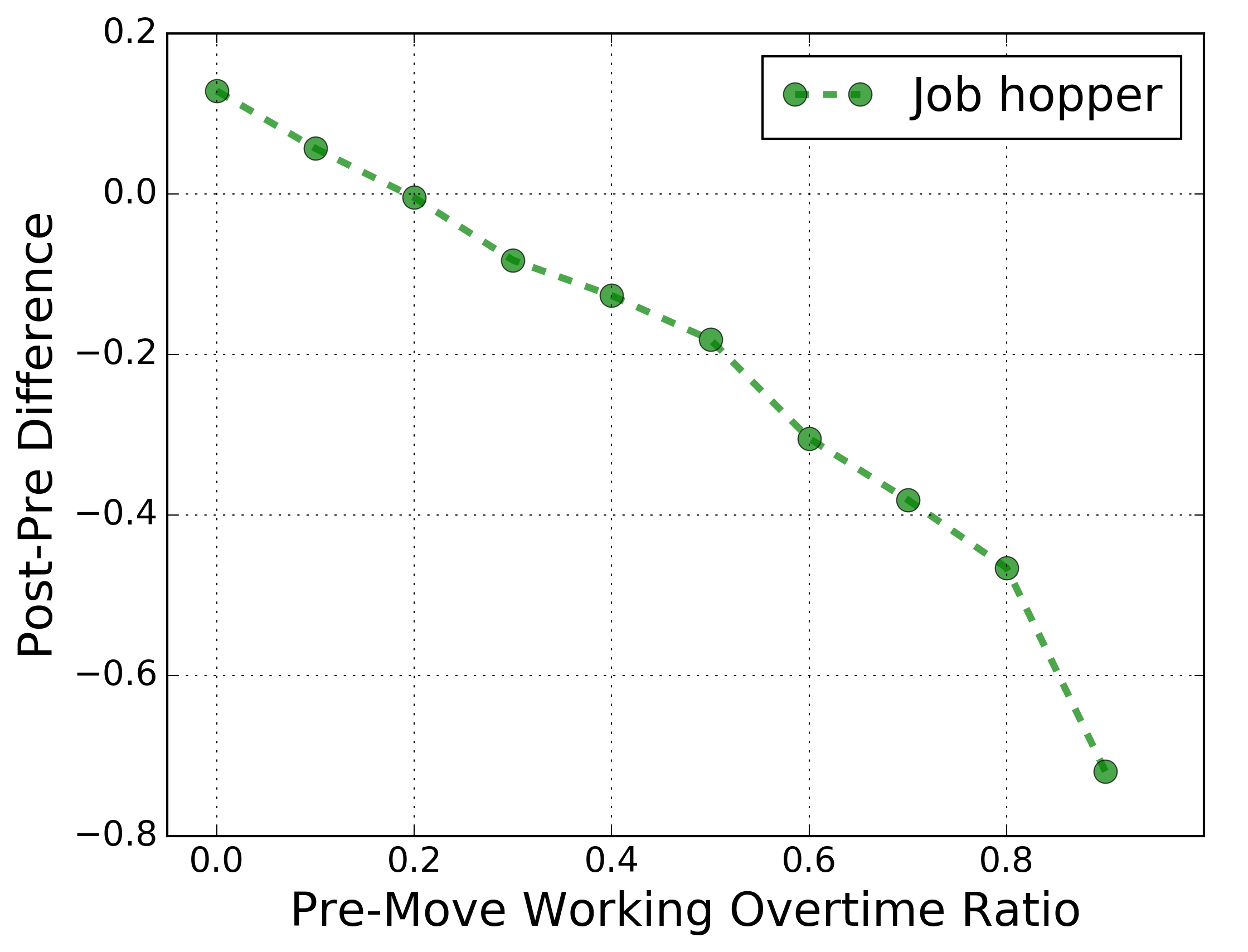}
		\caption{Post-Pre Difference}
		\label{fig:work_difference}
	\end{subfigure}
	\caption{a) Density plots of working overtime ratio for different user groups, b) Average changes at different pre-move working overtime ratio (which are divided into 0.1 bins).}
	\label{fig:work_time}
\end{figure}

\vspace{0.15cm}
\textit{\textbf{Observation 5:} People subject to higher housing cost are more likely to reduce their housing cost after moving. } 
\vspace{0.15cm}

The housing price embeds many geographic and economic factors, e.g., access to employment, amenities, schools. Within cities comparable housing units may be priced differently due to these factors. For example, Beijing Administrative Center's move to \textit{Tongzhou} district led to significant increase in housing price (and rent) in that region. Understanding whether people are upgrading or downgrading their housing can shed light on the health of local property markets and the economy as a whole. We collected monthly house transactions from different locations of Beijing and try to estimate the impacts of housing moves on housing expenses. Since we only estimate a user's home location, we cannot directly link a move to a transaction (or the rental of a new unit), so instead we match a move to the nearest spatial and temporal house transaction. It should be noted that although most migrants tend to rent instead of buying a permanent house, the rents are highly correlated with actual housing prices in Beijing. 

In total, we collected 10,140 transaction records (with average sale price) from 587 real estates located in the study area. For pre- and post-move $H$ hubs, we compute the median housing price in the same month/year from nearby real estates within \unit[3]{km} radius. The corresponding housing prices are searched separately for pre- and post-move $H$ hubs. The total number of matched housing moves is 1,727 (about 70.5\% of the detected housing moves). 

As shown in Table~\ref{tb:house_price}, on average, home movers have slightly higher housing price than stayers but the difference is not significant. Also, there is no significant difference between home movers' pre- and post-move hubs' housing price. However, when examining users with different pre-move housing price as shown in Figure~\ref{fig:house_difference}, we observe that users subject to higher pre-move housing price are more likely to reduce their housing cost after moving. 

\begin{table}[h]
	\centering
	\caption{Comparison on housing prices (Stayers: home hubs are used, Home movers: pre-move hubs are used). }
	\small
	\begin{tabular}{V||S|S}
		\hline
		\textbf{User Group} & \textbf{Avg Housing Price (RMB$/m^2$)}      & \textbf{t-statistic (Comparison vs. Stayer)}         \\ 
		\hline \hline 
		Stayers (home) & 53,201 & /  \\ \hline
		Home movers  & 53,787 & 0.958  \\ \hline		
		\addlinespace[0.75ex]
		\multicolumn{2}{l}{\textsuperscript{***}$:p<.001$, 
			\textsuperscript{**}$:p<0.01$, 
			\textsuperscript{*}$:p<0.1$}
	\end{tabular}
	\label{tb:house_price}
\end{table}

\begin{figure}[t]
	\begin{subfigure}[h]{0.235\textwidth}
		\centering
		\includegraphics[width=\textwidth]{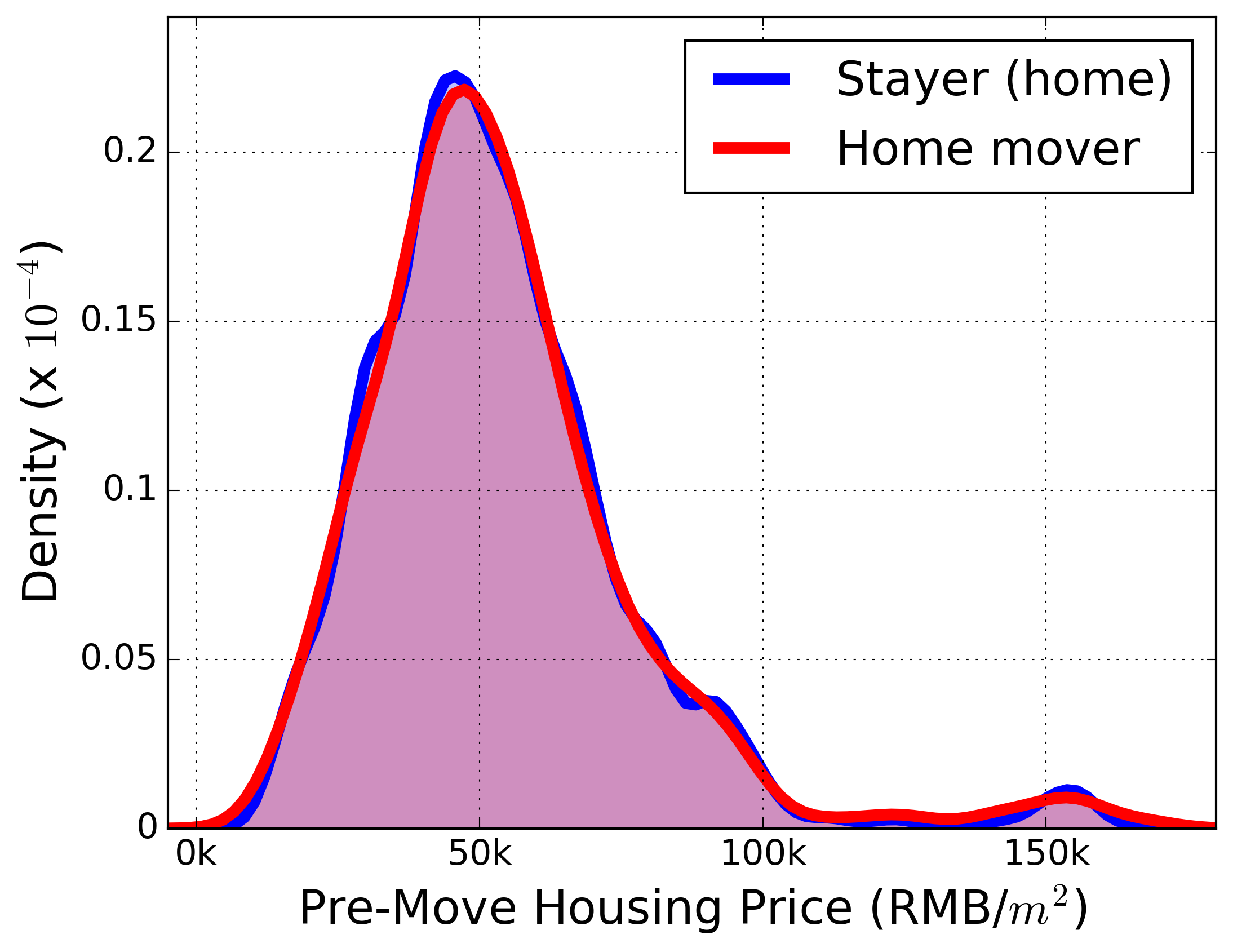}
		\caption{Density plot}
		\label{fig:house_density}
	\end{subfigure}
	\begin{subfigure}[h]{0.23\textwidth}
		\includegraphics[width=\textwidth]{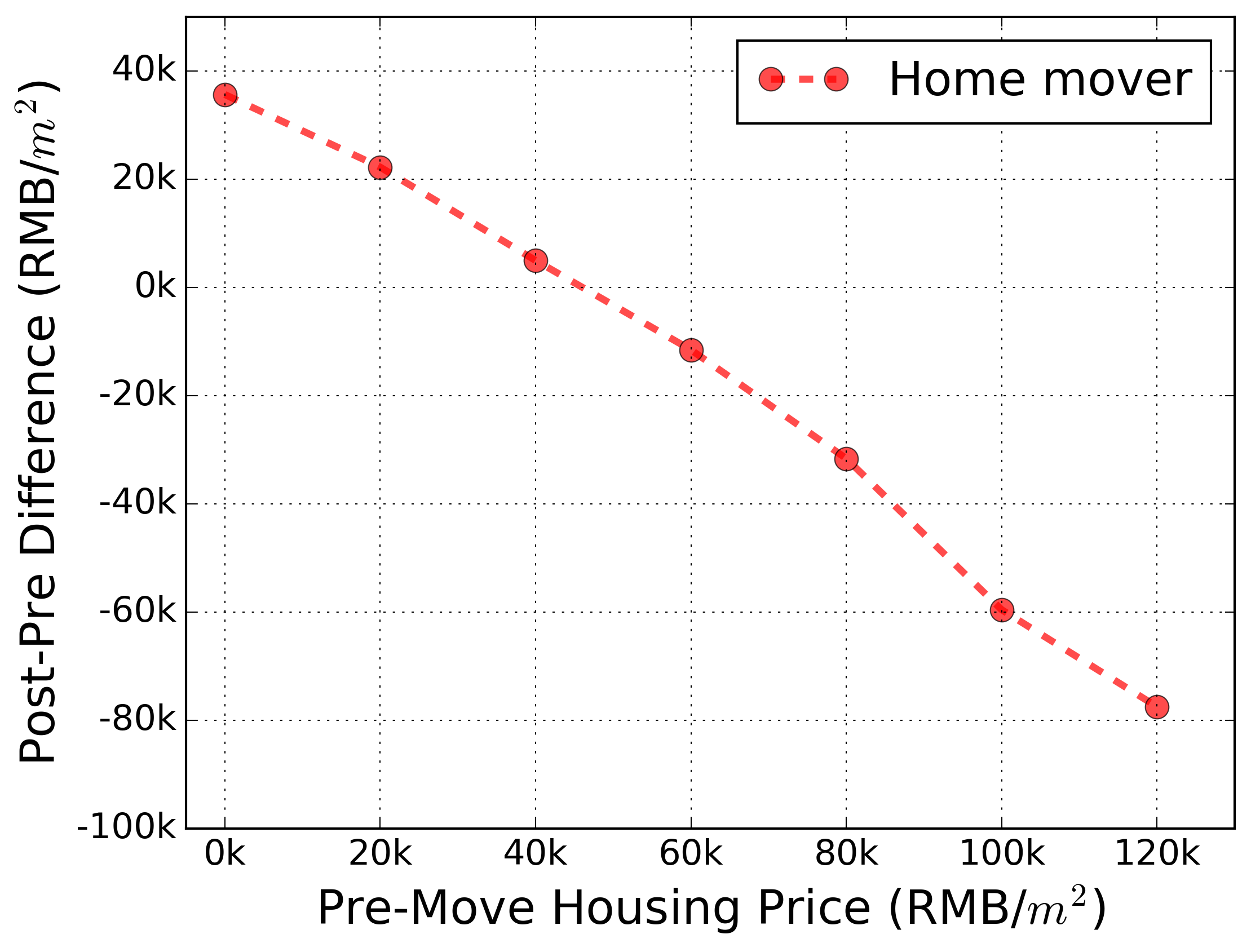}
		\caption{Post-Pre difference}
		\label{fig:house_difference}
	\end{subfigure}
	\caption{a) Density plots of housing prices for user groups, b) Average changes in housing prices at different pre-move housing price (which are divided into 20k RMB$/m^2$ bins).}
	\label{fig:house_price}
\end{figure}

\vspace{0.1cm}
\textit{\textbf{Observation 6:} There is a trade-off between lower housing cost and shorter commuting distance. } 
\vspace{0.15cm}

Beijing is a mega-city with multiple ring roads and the constructions of ring roads have witnessed Beijing's urbanization process and urban-rural integration. For instance, the construction of the 7th Ring Road was just completed in 2018. Therefore, we can roughly use ring roads to divide our study area into different regions as follows: \textit{city core} (within the 4th ring road), \textit{inner suburbs} (between 4th and 5th ring road) and \textit{outer suburbs} (between 5th and 6th ring road). 

In terms of housing move destinations, about 39.3\% were moving towards city core and about 60.7\% were moving towards inner or outer suburbs, which is consistent with the observation that most migrants reside outside central urban areas~\cite{liu2018urban}. The general trend is that suburbs were attracting more housing moves. As shown in Table~\ref{tb:mobility_direction}, people moving from city core to suburbs on average have a reduction in housing cost and an increase in commuting distance, and the farther they move out (i.e., to outer suburbs), the greater changes they achieve. These results demonstrate a trade-off between lower housing cost and shorter commuting distance for home movers in Beijing, and these two factors usually cannot be well satisfied at the same time. 

%
%
\begin{table}[h]
	\centering
	\caption{The Post-Pre differences in housing price and commuting distance for different housing move trajectories. }
	\begin{threeparttable}
		\small
		\begin{tabular}{c|c||V|V}
			\hline
			\textbf{From} & \textbf{To} & \textbf{Avg Difference in Housing Price (RMB/$m^2$)} & \textbf{Avg Difference in Commuting Distance (km)}     \\ 
			\hline \hline 
			City core  & City core & 1,675 & -0.76 \\ 
			&  Inner suburbs & -2,595 & 1.65 \\ 
			&  Outer suburbs & -13,661 & 3.53 \\ \hline
			Inner suburbs  &  City core & 6,965 & -0.73 \\ 
			&  Inner suburbs & -7,807 & -2.81 \\ 
			&  Outer suburbs & -6,302 & 3.33 \\ \hline
			Outer suburbs  & City core & 18,208 & -5.39 \\ 
			&  Inner suburbs & 6,922 & -5.19 \\ 
			&  Outer suburbs & -663 & 0.54 \\ \hline
		\end{tabular}
	\end{threeparttable}
	\label{tb:mobility_direction}
\end{table}
%

\section{Concluding Discussion}
\label{sec:conclusion} 

In this work, we explore the usage of a new modality of dataset, online food delivery dataset, to detect and characterize job and housing mobility. We design systematic methods to detect home and work ``hubs'' and use hub changes to study job and housing moves. The implications of this study are: First, information on job and housing moves is not available in the Chinese census data. Our work, leveraging food delivery service data, is able to provide analysis on higher temporal and spatial resolutions than most official sources. Second, we illustrate how novel forms of data could shed light on observing meso-scale urban movements, i.e., urban dynamics unfold over months or years. Third, we examine a series of macro- and micro-factors and link them with individual moves, the observations align with widely observed trends and they also provide ``soft'' validations on the dataset and the methods we propose. The results of this study demonstrate the effectiveness of using this new delivery-based data to analyze job and housing mobility. The major limitation of this study is data representativeness. We are very cautious about this and therefore use various data sources, e.g., census, previous literatures, to examine the results derived. The study of migrations using new forms of data is a challenging while promising field. It might also be used to improve both the design and estimates from representative surveys~\cite{spielman2017potential}. As future work, we plan to develop ensemble approach to utilize ``organic'' data, for instance, using multiple data modalities that jointly capture job and housing dynamics from different perspectives. 

\bibliographystyle{IEEEtran}
\bibliography{main}

\end{document}